# Augmenting Italian Cultural Heritage with Virtual and Digital Technologies: the final outcomes of Project CHANGES' Spoke 4

Silvio Peroni[1,*], Gianluca Genovese[2], Roberto Balzani[1], Silvano Montaldo[3], Sofia Pescarin[4], Cristina Caterina Amitrano[3], Luisa Ammirati[1], Riccardo Antonino[5], Giorgio Bacci[6], Davide Bagnaresi[1], Sebastian Barzaghi[1], Giuliana Benvenuti[1], Marco Bertini[6], Luca Bevilacqua[7], Marco Biffi[6], Elisa Bonacini[8], Federica Bonifazi[4], Alice Bordignon[1], Davide Borra[5], Andrea Bottino[9], Ivana Bruno[10], Daniele Caccavale[2], Marcello Calogero[1], Giuseppe Capotorto[11], Andrea Carpentieri[12], Veronica Casadei[1], Cristina Casero[13], Maurizio Cattani[1], Roberto Cavallo Perin[3], Vito Leonardo Chiechi[14], Fabio Ciotti[15], Luca Cipriani[1], Simona Colitti[1], Federica Collina[1], Michela Contessi[1], Lorenzo Copelli[13], Michele Corriero[8], Marianna Cuomo[2], Francesca D'Angelo[1], Rossana Damiano[3], Marilena Daquino[1], Stefano De Martino[3], Kevin De Vecchis[6], Davide Domenici[1], Nicole Dore[16], Francesca Fabbri[1], Bruno Fanini[4], Filippo Fantini[1], Fabrizio Federici[6], Daniele Ferdani[4], Enrico Ferraris[17], Giulia Fiorini[1], Anna Forte[1], Rocco Furferi[6], Riccardo Gagliarducci[18], Vincenzo Gattulli[110], Manuel Gentile[4], Federica Giacomini[1], Valentina Alena Girelli[1], Martina Grinello[1], Bianca Gualandi[1], Jasmine Habsy[13], Stephan Hassam[20], Ivan Heibi[1], Alessandro Iannucci[1], Alessandra Inglese[15], Marco Lamorte[1], Laura Travaglini[4], Federica Maietti[21], Rachele Manganelli Del Fà[4], Gianluigi Mangiapane[3], Nicola Mariniello[7], Giulia Marsili[1], Arcangelo Massari[1], Marcello Massidda[4], Michele Mellara[22], Enrico Mensa[3], Mattia Modugno[16], Roberto Montanari[2], Antonio Monteleone[16], Arianna Moretti[1], Elena Musiani[1], Francesca Nicolais[2], Sara Obbiso[1], Carmelo Occhipinti[15], Francesca Paruzzo[3], Cecilia Pennacini[3], Davide Perinciolo[23], Mario Petrella[1], Margherita Elena Pomero[1], Roberta Presta[2], Maria Felicia Rega[4], Giulia Renda[1], Diego Ronchi[4], Alessandro Rossi[22], Nicola Santopuoli[1], Chiara Scardozzi[1], Nicolò Serafino[13], Ilaria Serati[3], Ulderico Sicilia[24], Stefano Sorrentino[1], Daniele Spedicati[1], Cristina Stalteri[3], Riccardo Stracuzzi[1], Mattia Sullini[1], Ursula Thun Hohenstein[21], Maria Alessandra Tini[1], Federica Veratelli[13], Riccardo Zanzi[25], Guido Nicolas Zingari[3]

[1] University of Bologna, Italy
[2] University of Suor Orsola Benincasa, Italy
[3] University of Turin, Italy
[4] National Research Council, Italy
[5] Robin Studio, Italy

[6] University of Florence, Italy
[7] Engineering I.I., Italy
[8] University of Bari, Italy
[9] Polytechnic of Turin, Italy
[10] University of Cassino and Southern Lazio, Italy
[11] DTC Lazio, Italy
[12] Imago, Italy
[13] University of Parma, Italy
[14] Digitarca, Italy
[15] University of Rome Tor Vergata, Italy
[16] NAIS, Italy
[17] Museo Egizio, Italy
[18] Brixel, Italy
[110] Sapienza University of Rome, Italy
[20] Randolph-Macon College, USA
[21] University of Ferrara, Italy
[22] Mammut Film, Italy
[23] Ribes DigiLab, Italy
[24] Risviel, Italy
[25] TP Design, Italy

[*] Corresponding author: Silvio Peroni, silvio.peroni@unibo.it, Department of Classical Philology and Italian Studies, Via Zamboni 32, 40136 Bologna (BO), Italy.

# Abstract

CHANGES (Cultural Heritage Innovation for Next-Gen Sustainable Society) was a project coordinated by the CHANGES Foundation, bringing together complementary disciplines and expertise across the entire cultural heritage lifecycle. This article focuses on the project's research area dedicated to applying virtual technologies to museums and art collections. This research enabled us to experiment with different types of museums and art collections across Italy, designing case studies and best practices that institutions and similar contexts can adapt and reuse. The meta-analysis of the case studies showed that museum types play a decisive role in shaping not only technological choices but also organisational strategies, training models, and pathways to sustainability. This work also allowed us to identify technological and organisational gaps that go beyond single cases and reflect shared challenges across museum contexts. Finally, we documented the success of the engagement, exploitation, and dissemination strategy proposed in the project.

# Introduction

Following the outbreak of the COVID-19 pandemic in 2020, the European Commission allocated substantial funds (i.e. 750B euros) to support several areas, including governance, infrastructure, civil society, and research. Of this amount, Italy has received 194,4B euros under its National Recovery and Resilience Plan (Piano Nazionale Di Ripresa e Resilienza, 2021). Of these, 30,09B euros have been made available to support excellence in research, which has been implemented through calls organised in several macro areas. Among these, one was dedicated to the creation of *enlarged partnerships* between universities, research centres, and companies, which have been assigned an overall amount of 1,61B euros to develop innovative research projects on various fields including sustainable mobility, alternative energies, superconductors, prevention and monitoring of climate change, and cultural heritage.

One of these initiatives is CHANGES (Cultural Heritage Innovation for Next-Gen Sustainable Society), the only Extended Partnership specifically devoted to cultural heritage. Such a project is coordinated by the CHANGES Foundation (https://www.fondazionechanges.org/en/spoke-4-en/), a non-profit organisation established in September 2022 to serve as its central Hub following the call issued by the Italian Ministry of University and Research (Avviso n. 341 Del 15-03-2022, 2022). In this capacity, the Foundation coordinates the research activities of the project's partners and Spokes, bringing together complementary disciplines and expertise across the entire cultural heritage lifecycle—from collecting and processing multidisciplinary data to phygital or virtual visualisation and enhancement for tourism purposes.

The work presented in this article was carried out within CHANGES between December 2022 and April 2026 and, more specifically, within Spoke 4, the project's research area dedicated to the application of virtual technologies to museums and art collections. Indeed, in the past years, we have observed the proposition of several national and international policies aiming at supporting the universal use of digital data of the cultural heritage domain. Some of these initiatives – such as those by UNESCO (2009), the European Union (Commission Recommendation of 27 October 2011 on the Digitisation and Online Accessibility of Cultural Material and Digital Preservation 2011), and the Italian Ministry of Culture (Aprea, Diego et al. 2023) – have provided either support or specific guidelines for cultural heritage digitisation in the past years. To comply with these policies and guidelines, we need to make the digital enhancement of cultural heritage a permanent and widespread practice in cultural heritage institutions to increase the knowledge, curation and management of artefacts in all forms, expand the involvement of the general public, thus, improving accessibility, inclusiveness, critical thinking, participation, enjoyment and sustainability.

Within this international context, the primary goal of CHANGES' Spoke 4 was to focus on digital cultural heritage (DCH)'s impact, comparing it with the current view on tangible and intangible heritage. DCH objects are defined through the network of interlinked relations they have with the cultural heritage environment and their provenance context, while (in)tangible objects result from selective processes defined and used by cultural heritage institutions over time.

Among the various research questions introduced in the project, some of them have been addressed in already published works such as (Barzaghi et al., 2024a, 2024b; Barzaghi et al. 2025a, 2026); therefore, the dimensions we discuss in this article concern the aspects related to the case studies and implementations finalised recently:

1. How do we harmonise the different heritagisation and value creation processes behind DCH and (in)tangible heritage?
2. How do we preserve, exploit and make cultural heritage accessible, specifically the potential experience a visitor can have with it?

To address these questions, the Spoke has experimented with different types of museums and art collections across Italy to design pilot studies and best practices that can be adapted and reused in institutions and contexts with similar characteristics. These templates (Types A-F) are listed as follows:

- Type A – high-density and innovative museum;
- Type B – natural history and scientific museum;
- Type C – widespread art gallery;
- Type D – sites museums with tangible and intangible heritage and landscapes, among those that have either low or no virtual and digital implementations with a particular focus in southern Italy;
- Type E – historical palaces, labelled as museums, among the most important in the Italian heritage context, including UNESCO sites;
- Type F – demo-ethnic-anthropological museums, including rural culture museums (with less than 20,000 visitors per year), anthropological museums and museums of art and popular traditions.

Following these templates, we designed and ran several studies to address the RQs presented above. Specifically, we examined the relationship between the initial and final cultural knowledge conveyed by Digital Cultural Heritage (DCH) objects, addressing the interlinking between them (i.e. the relational heritage) and (in)tangible cultural heritage represented in the current narrative in museums and art collections. We have contacted existing cultural heritage institutions that were compliant with the six templates (Types A-F) mentioned above to set up collaborative research projects, called *core case studies*, that have been supported by funds from Spoke 4's Cascade Calls – i.e. a formal application to involve external parties (companies, other institutions, research centres, etc.) that worked together with Spoke 4 researchers – to enable their implementation, to improve and augment the valorisation of their cultural heritage.

To this end, we have experimented with and reused a plethora of virtual (energy efficient) technologies, which include:

- decentralised and interlinked knowledge graphs of DCH and (in)tangible objects (including provenance information);
- Web-based environments for sharing cultural heritage and involving users in museums and art collections in situ or remotely;
- new design approaches to virtual technologies for cultural heritage including eXtended Reality (XR, i.e. virtual reality, augmented reality, mixed reality, immersive reality), gamification, serious games, edutainment;
- 2D/3D models and multimedia (including video storytelling);
- tools for digitisation and simulation for enabling digital approaches to cultural heritage, Internet of Things and sensor networks;
- AI-based methods and tools for cultural heritage;
- location-based technologies connected to GIS for cultural heritage (e.g. sites).

Working in strict cooperation with cultural heritage institutions has enabled us to define and share guidelines for instructing institutions and researchers on the processes and requirements to set up appropriate workflows to acquire and digitise cultural heritage in a way that is compliant with European policies on Open Science (Council of the European Union 2025; European Commission, Directorate General for Research and Innovation 2021), and to develop a set of open tools and data that can also be reused by other national institutions that were not involved directly in the research of the Spoke, to maximise the investments of Project CHANGES and to enable future users to adopt these technologies after the project finish. These core case studies were also accompanied by additional *research case studies* compliant with Types A-F, which involved Spoke 4's partners (without direct support from the Cascade Calls) to experiment with further uses of the aforementioned guidelines and tools within the project's time span.[1]

The remainder of the article is structured as follows. In Section "Methodology", we present the methodology we have set up for running the entire Spoke 4, introducing also the approach used for assessing the results obtained during the project. In Section "Stakeholders identification", we describe the approach used to identify the relevant stakeholders to involve in the project to implement targeted exploitation actions to disseminate project results. In Section “Case studies”, we briefly describe all the case studies conducted involving cultural heritage institutions and companies, including those implemented with the support of Spoke 4’s Cascade Calls (core case studies) and those conducted by Spoke 4 researchers without direct support of these additional funds. In Section “Results”, we present the main outputs of the research conducted by Spoke 4. Finally, Section “Discussions and conclusions” concludes the article and outlines future developments.

---

[1] A more detailed and technical description of parts of the tools and results we introduce in this article can be found in the Spoke 4 deliverables, listed in Appendix 1.

# Project organisation and evaluation methodology

Spoke 4 has been organised into two distinct temporal frames: the first (*preparation phase*) focused on preparing the setting for the research, while the second (*execution phase*) focused on implementing actions and conducting assessment exercises. They are briefly detailed as follows.

**Preparation phase.** In this phase, the main aim was to meet with several cultural heritage institutions across Italy to define suitable shared research lines and to co-design experimental settings that enable research to advance in the application of digital and virtual technologies in the CH context. At the same time, we aimed to align the proposed initiatives with next year’s plans of the cultural heritage institution involved, aiming at having a mutually beneficial collaboration. In addition, we wanted to collect partners' interest and proposals for additional experimental research studies that would not have the support of the Cascade Call. Furthermore, we aimed to set up an additional *pilot case study* based on an exhibition that included several heterogeneous cultural heritage objects. In our intentions, this pilot study serves two main purposes: to test the use of virtual technologies that should be used in the context of the “core” case studies, and to derive the guidelines to share with the companies and institutions funded through the Cascade Calls to implement the research activities of the “core” case studies. In this context, an important effort is also devoted to identifying potential stakeholders interested in implementing the research in the “core” case studies, to involve them in future exploitation activities.

**Execution phase.** This phase had two main objectives. The first was to finalise the pilot case study to showcase a clear example of all the technologies used and developed in the context of Spoke 4 for the valorisation of cultural heritage. The second was to publish the guidelines for running the whole acquisition and digitalisation process for cultural heritage objects and publishing them in a FAIR-compliant (Wilkinson et al. 2016) way following the Open Science principles (UNESCO 2021). These guidelines were intended to support researchers involved in implementing the “core” case studies and research studies. In addition, we wanted to develop and share prototypes to be used in the context of the “core” case studies as support for all the companies and groups that would win the Cascade Calls to be published during this phase. Finally, we wanted to implement tools to monitor and assess both the case studies and the involvement and exploitation of all the research products and expertise created during the project.

To implement the preparation and execution phases, we organised the work into five work packages (WPs), each addressing specific aspects of the overall research.

**WP1 – FAIR Data and Spoke Coordination.** This WP ensures compliance with FAIR principles for all data produced – including digital rights management and ethical issues – and regulates day-by-day operations for monitoring and coordinating the Spoke to reach technical and scientific goals on time, setting up appropriate channels for document exchanges and communication between all the partners.

**WP2 – Cultural Context.** This WP analyses the status of museums and art collections in national and international contexts, identifying existing technology-aided museum narratives and how virtual technologies can expand the current space-time perception of (in)tangible cultural heritage, and highlighting functional requirements and constraints linked to case studies. Particular attention is given to studying and developing guidelines for enabling virtual technologies to go beyond physical constraints and expand knowledge and narratives of DCH and (in)tangible objects.

**WP3 – Technological Context.** This WP studies existing virtual technologies for use in the museums and art collections analysed by WP2, whether in their final state (i.e. without increasing collections capabilities) or under development, to identify the most appropriate ones for the pilot studies (WP4). Appropriate prototypes are developed when necessary to implement the guidelines/best practices introduced in WP2.

**WP4 – Pilot Studies.** This WP uses the guidelines and prototypes identified in WP2 and WP3 in several concrete pilot studies covering all identified museum and art collection templates (Types A-F). Evaluations of both guidelines and prototypes are performed with real users to track the effectiveness and appropriateness of the solutions proposed in each pilot and to identify strengths and weaknesses.

**WP5 – Involvement and Exploitation.** This WP aims at maximising the spoke impact towards the research and cultural heritage communities and society by developing specific communication strategies for supporting the Spoke to scale the technologies up at a national level, fostering their adoption in other cultural institutions and developing an ecosystem of services to support citizens and cultural heritage operations by creating cultural offerings as well as new job profiles.

In addition to this organisational framework, we have also developed methods for assessing the various dimensions and outcomes of the research conducted. The following subsections briefly introduce these methods.

## Analysis of the case studies

We set up an assessment method that considers a qualitative cross-case synthesis approach based on the systematic analysis of documentary sources. The primary data come from the “Guide to Implementation” documents produced for each case study (both “core” and research), following a shared structure (Amitrano et al. 2025) that covers objectives, technological choices, organisational processes, experimentation, unforeseen issues and future lines of research. Rather than applying line-by-line coding of textual data, each case study was treated as a unit of analysis and assessed against a predefined *analytical framework*. This approach is particularly suited to document-based meta-analysis, where the aim is to compare implementation processes and outcomes across cases using shared dimensions of analysis.

The analytical framework was co-developed with the Spoke 4 team and allowed us to create a framework suitable for all museum types (A-F), reflecting their peculiarities and specificities. In particular, such a framework includes dimensions related to:

- contextual characteristics and museum typology;
- technological solutions;
- digitalisation processes and user experience design;
- knowledge sharing, training, and organisational adoption;
- experimentation with users and evaluation methods;
- critical issues, technological gaps, and future research needs.

This framework would be applied consistently across all case studies and would serve as the reference structure for data extraction, comparison, and synthesis. The results of the individual case assessments would then be consolidated into the comparative analytical matrix shown in Table 1, which serves as the core instrument for cross-case synthesis. Consistently with the research design, the matrix does not aim to provide a purely descriptive overview of the cases, but rather to support a structured comparison of implementation processes across heterogeneous museum contexts.

| **Dimension of analysis** | **Description** | **Analytical purpose** |
|---|---|---|
| Type | Classification of the cases according to museum types A-F | To contextualise technological and organisational choices |
| Case study | Core case study / Research case study | To account for differences in implementation maturity |
| Technological role | Main function of technology (e.g., immersive experience, curatorial infrastructure, etc.) | To relate technology to institutional needs |
| Technological configuration | Main technological components and architectural approach | To compare complexity, modularity, and interoperability |
| Familiarisation model | Training models, workshops, manuals, co-design activities | To analyse adoption strategies and capacity building |
| Organisational embedding | Degree to which the technology is integrated into daily practices | To evaluate sustainability beyond the project |
| Experimentation status | Planned, ongoing, completed, or not comparable | To frame results without forcing equivalence |
| Key challenges | Technical, organisational, infrastructural, legal, or ethical issues | To identify recurring barriers |

**Table 1.** Comparative analytical matrix for case studies.

Each case study is treated as a distinct unit of analysis and positioned in the matrix of Table 1 according to a shared set of analytical dimensions derived from Amitrano et al. (2025). These dimensions reflect the key aspects investigated across the cases, including museum typology, technological configuration, alignment between institutional context and technological choices, approaches to staff and stakeholder familiarisation, degree of organisational embedding, and readiness for experimentation and evaluation.

The matrix enables comparison at multiple levels. First, it allows cases within the same museum typology to be examined together, highlighting typology-specific implementation patterns and recurring challenges. Second, it supports cross-typology comparison, making it possible to identify how similar technological approaches are adapted to different institutional, spatial, and organisational conditions. Finally, by explicitly distinguishing between "core" and research case studies, the matrix accounts for differences in implementation maturity and scope, preventing inappropriate direct comparisons between cases with different objectives and levels of deployment.

Through this structured alignment, the matrix helps identify recurring patterns, convergences, and divergences across the case studies. In particular, it supports abstracting relationships between museum typologies and technological choices, as well as between technological complexity and strategies for organisational adoption. At the same time, it allows cross-cutting issues (e.g. sustainability, governance, capacity building, and infrastructural constraints) to emerge across otherwise heterogeneous contexts.

## Tracking involvement and exploitation

In the context of the plan for supporting the involvement and exploitation of the outcomes of Spoke 4, the main objectives we wanted to measure concern (a) the development of *exploitation* strategies for maximising reusability and enabling the scaling up of narratives and prototypes at the national level, and (b) the creation of a workflow to improve the cultural value of cultural heritage objects through digital *dissemination* approaches (e.g. targeting society, high schools, and museum networks), assessing their impact on potential audiences.

Concerning objective (a), two key performance indicators (KPIs) are used, i.e. the quantitative size of the community involved (QNC) and the qualitative response of the community involved (QLC), which represent the inherent tension between, on the one hand, the breadth and representativeness of the engagement process (QNC) and, on the other hand, the soundness and coherence of the strategic development process (QLC).

QNC is operationalised as a set of quantitative and process-oriented indicators reflecting the scale, diversity, and structuring of stakeholder engagement activities supporting the development of the Spoke 4 exploitation plan:

- *number of stakeholders involved in the development of the exploitation plan*, measured through the number of participants in workshops designed and run for this purpose;
- *distribution of involved stakeholders in the exploitation plan design* across the quadruple helix categories (academia, government, industry, civil society), used to assess the representativeness of the engagement process (Carayannis and Campbell 2009);
- *number of stakeholder engagement activities* implemented in support of exploitation strategy development, considering engagement activities in relation to the "core" case studies, which represent the main application contexts of Spoke 4;
- *distribution of stakeholder engagement activities* across the quadruple helix categories, used to verify that engagement actions were not concentrated on a single stakeholder group but distributed across different perspectives relevant to exploitation and scaling up.

The ultimate success of the exploitation strategies is expected to be reflected in the reuse and re-adoption of the developed prototypes, workflows, and solutions in comparable cultural heritage initiatives. However, the reuse and re-adoption of Spoke 4 outcomes cannot be fully measured within the current project timeframe, since these are inherently long-term effects. For this reason, and because the extent of reuse and re-adoption is intrinsically linked to the perceived value and appreciation of the developed solutions, we adopted a proxy strategy to preliminarily address the QLC dimension by focusing on the involvement of civil society and targeted stakeholders through dedicated user testing activities. While not equivalent to actual reuse, these activities provide early qualitative evidence on the acceptability, perceived usefulness, and potential readiness for adoption of the developed solutions, informing exploitation considerations. From the perspective of technology acceptance theories, and in particular the Technology Acceptance Model (TAM) (Davis and Granić 2024), such dimensions represent well-established antecedents of technology diffusion and adoption. Perceived usefulness and perceived ease of use (Davis 1989) are recognised as key factors influencing users' intention to adopt a technological solution, which in turn is a prerequisite for its wider spread and reuse.

Thus, considering QLC, we provide specific guidelines for organising user testing sessions across the case studies to gather appropriate feedback through established and validated evaluation instruments, targeting dimensions relevant to future reuse and adoption. In particular, this approach is operationalised through the following indicators:

- *number of users* involved in the testing activities;
- *percentage* of case studies for which user testing was conducted, ensuring that qualitative evidence was collected across a representative subset of the project's application contexts;
- *usability*, assessed through the System Usability Scale (SUS) (Brooke 1996), where applicable;
- *overall user experience quality*, evaluated using the Short User Experience Questionnaire (UEQ-S) (Schrepp et al. 2017), where applicable.

- *endorsement*, measured through the Net Promoter Score (NPS) (Reichheld 2003), where applicable.

Usability, user experience, and endorsement are defined a priori as part of the user testing framework to adopt for the case studies. For each dimension, we suggested one reference instrument. To support their consistent and informed use, a dedicated set of guidelines for UX evaluation would be shared with all project partners, allowing participants to select the tools most appropriate to their specific testing context.

The recommended UX assessment instruments are introduced as qualitative monitoring tools. Their primary function is not to determine compliance with preset project performance benchmarks, but to systematically document and track how the developed solutions are perceived in terms of usability, experience quality, and endorsement within each use-case context. Given the exploratory nature of the project and the diversity in technological maturity levels, user groups, and contextual constraints across the core cases, imposing uniform numerical thresholds would have risked oversimplifying highly differentiated validation environments. The observed values therefore serve as an evidence-based baseline for future benchmark definition, enabling longitudinal tracking within each use case and supporting more structured target setting in subsequent project cycles or scaling initiatives.

Concerning objective (b), which focuses on the creation of a workflow to improve the cultural value of cultural heritage objects through digital dissemination approaches including the preliminary assessment of their impact on potential audiences, each "core" case study represents an applied example of how cultural heritage objects are translated into digital dissemination solutions, following a consistent set of methodological choices related to content selection, interpretation, digital mediation, and audience targeting. Taken together, these case studies exemplify the workflow and demonstrate its applicability across different cultural contexts and dissemination scenarios.

For this objective, we rely only on QLC to assess the workflow's impact on potential audiences, using the same qualitative indicators adopted for objective (a). Indeed, these measures provide early qualitative evidence of how effectively the dissemination approaches implemented in the core use cases support meaningful interaction with cultural content, thereby informing the assessment of the workflow's potential impact on its intended audiences.

Finally, to measure the effectiveness and impact of the involvement, exploitation, and communication activities carried out, a set of additional KPIs is identified:

- for the exploitation, the preparation of at least 6 publications and the participation in at least 3 conferences/events for scientific dissemination in three years;
- for communication, the publication of 1 post on social media per month, and the publication of 1 article for the newsletter every two issues;
- for stakeholders' involvement, identify at least 10-15 stakeholder categories to be involved in Spoke 4, plus identify at least 1 professional network.

# Stakeholders identification

As anticipated in the previous section, the involvement of stakeholders represents a fundamental pillar of Spoke 4 and is driven by three main objectives: (a) identifying and engaging key stakeholder categories, (b) implementing targeted exploitation actions to disseminate project results, and (c) designing and deploying effective communication materials and tools to reach diverse audiences and maximise the visibility of the project and its outcomes. As part of the actions to ensure effective stakeholder involvement, particular emphasis is placed on the participatory approach adopted to map stakeholders and define engagement strategies. This process includes identifying and clustering stakeholders based on their interests and alignment with project objectives, defining their roles, and monitoring engagement activities through dedicated KPIs.

To implement these objectives and foster collaboration among project partners, we designed and facilitated the workshop, entitled “CultureInDialogue”, which served as a key moment for co-creating a shared stakeholder engagement model and identifying strategic approaches to enhance the dissemination and sustainability of Spoke 4’s research outputs. The workshop model enabled close interaction with members identified by each Spoke 4 participating organisation by sharing the process of mapping key stakeholders and stakeholder networks, as well as the most effective ways to engage them. The workshop was grounded in the quadruple helix model, a well-established framework for fostering innovation in the knowledge economy through collaborative interaction among Government, Industry, Academia, and Civil Society (Carayannis and Campbell 2009). Drawing on this model, the workshop pursued three main objectives:

1. *issue mapping* – concerning the identification of key stakeholders in the cultural heritage sector and their relevant networks within industrial, academic, institutional, and civil society domains, clustering them according to their specific interests and expertise related to cultural heritage and the scope of Spoke 4 activities.
2. *influence mapping* – concerning the evaluation of stakeholder influence in the cultural heritage landscape, assessing their potential impact on Spoke 4 goals, and the definition of a strategic framework for engagement based on the relative influence and positioning of each stakeholder or network;
3. *involvement and engagement strategy* – concerning the assessment of the current level of involvement of each stakeholder in ongoing Spoke 4 activities, developing tailored engagement strategies for each stakeholder or stakeholder group, ensuring alignment with the broader objectives of Spoke 4.

The following methods were implemented to ensure an optimal process of co-creation of strategies:

- conducting the workshop virtually to ensure maximum participation from partners;
- using virtual platforms for the collection of preliminary data functional for the participatory phases of the workshop (Google Modules, https://forms.google.com);
- using the MIRO (https://miro.com) visual co-design platform made it possible to work simultaneously by leveraging tools and templates to build the infrastructure needed to carry out the workshop phases, optimising processes without compromising the quality of the analysis, and indeed adding value on the methodological level and the results achieved.

The workshop has involved 23 active participants from Spoke 4 institutions, plus the organisers and moderators. A report that outlines the full set of activities conducted during the workshop and, based on the outcomes of collaborative sessions, provides a comprehensive overview of the stakeholder mapping and engagement strategy has been published in (Genovese and Montanari 2024), which also details the workshop structure, participant distribution, preliminary preparation, the three main phases of the workshop, and the outcomes of each phase. Table 2 synthesises the outcomes of the stakeholder mapping and engagement strategy co-designed during the workshop. As part of the participatory approach adopted to enhance stakeholder involvement, Table 2 reflects the structured process used to classify stakeholders according to the quadruple helix model and to assess their current level of engagement with Spoke 4 activities.

| **Stakeholder Segment** | **High Interest High Influence** | **Low Interest High Influence** | **High Interest Low Influence** | **Low Interest Low Influence** |
|---|---|---|---|---|
| Government<br><br>*19 stakeholders identified* | Direct connections with informed stakeholders<br><br>Face-to-face meetings<br><br>Consultative engagement aligned with policy needs<br><br>Risk mitigation of disengagement | Tailored and direct strategies<br><br>Invitations to relevant events (e.g., demonstrator openings)<br><br>Participation in targeted focus groups | Thematic consultations<br><br>Engagement through educational initiatives | Outreach strategies |
| Industry<br><br>*11 stakeholders identified* | Participation in trade fairs<br><br>Consultative focus groups<br><br>Updates on innovations and project activities | Focus groups to assess specific impacts<br><br>Customized engagement strategies | Training and informative content<br><br>YouTube tutorials and technical/scientific videos<br><br>Video content on BB.CC. topics<br><br>Scientific papers | Tailored and direct strategies |
| Academia<br><br>*20 stakeholders identified* | Advanced topic-specific tutorials<br><br>Engagement via LinkedIn<br><br>Informative YouTube content<br><br>Participation in academic fairs (e.g. Digital Heritage, Turisma) | Thematic focus groups<br><br>Virtual exhibitions<br><br>Peer-reviewed publications | Access to specialized training<br><br>Educational webinars and resources<br><br>Public-oriented presentations via online/offline channels | General awareness activities |

| Civil Society<br><br>*18 stakeholders identified* | Social media (e.g. Instagram)<br><br>Smartphone video games for cultural awareness<br><br>Events focused on cultural heritage | Informative approach for stakeholders such as banking foundations<br><br>Invitations to demonstrators and digital exhibitions<br><br>Sharing educational materials | Accessible digital content<br><br>Informal learning tools | Awareness strategies |
|---|---|---|---|---|

**Table 2.** A synthesis of the outcomes from the stakeholder mapping and engagement strategy co-designed during the workshop, used to assess current level of engagement of the stakeholders with Spoke 4 activities.

| GOVERNMENT | INFLUENCE AND INTEREST RATE | | | |
|---|---|---|---|---|
| INFORMATION RATE | HIGH INFLUENCE, HIGH INTEREST | HIGH INFLUENCE, LOW INTEREST | LOW INFLUENCE, HIGH INTEREST | LOW INFLUENCE, LOW INTEREST |
| FULLY INFORMED | Municipalities where Spoke 4 use cases are located. | Assessorato alla Cultura e al Turismo, Regione Piemonte | Cultural foundations | Uffici scolastici regionali |
| | | | Associazione Parchi letterari | |
| PARTIALLY INFORMED | Direzione Generale Musei del Ministero | Regioni | Digital Library | Conferenza dei Rettori delle Università Italiane (CRUI) |
| | Aziende di promozione turistica della Regione Basilicata. | | | |
| NOT INFORMED | Associazione Nazionale Musei Scientifici (ANMS) | | | Associazione Nazionale Comuni italiani (ANCI) |
| | Associazioni museali (AMACI, ICOM) | | | UNICEF |

| INDUSTRY | INFLUENCE AND INTEREST RATE | | | |
|---|---|---|---|---|
| INFORMATION RATE | HIGH INFLUENCE, HIGH INTEREST | HIGH INFLUENCE, LOW INTEREST | LOW INFLUENCE, HIGH INTEREST | LOW INFLUENCE, LOW INTEREST |
| FULLY INFORMED | Engineering | | | |
| | DATABENC (Distretto ad Alta TecnologiA per i BENi Culturali) | | | |
| | TICHE | | | |
| | ETT | | | |
| | Consorzio cultura e innovazione | | RE:LAB | |
| PARTIALLY INFORMED | | | Brixel | |
| | | | QAcademy | |
| | | | Alicubi | |
| NOT INFORMED | | | | |

| CIVIL SOCIETY | INFLUENCE AND INTEREST RATE | | | |
|---|---|---|---|---|
| INFORMATION RATE | HIGH INFLUENCE, HIGH INTEREST | HIGH INFLUENCE, LOW INTEREST | LOW INFLUENCE, HIGH INTEREST | LOW INFLUENCE, LOW INTEREST |
| FULLY INFORMED | Territorio di Aliano | | Associazione UCIIM | |
| | | | Territorio di Galtellì | |
| PARTIALLY INFORMED | Biblioteche, Archivi, Musei, Gallerie | | | |
| | | | | |
| NOT INFORMED | Fondazioni bancarie | | | Fondazione di Sardegna |
| | | | | Fondazione Piemonte dal vivo |
| | | | | Associazione categoria videoludica (IIDEA) |

| ACADEMIA | INFLUENCE AND INTEREST RATE | | | |
|---|---|---|---|---|
| INFORMATION RATE | HIGH INFLUENCE, HIGH INTEREST | HIGH INFLUENCE, LOW INTEREST | LOW INFLUENCE, HIGH INTEREST | LOW INFLUENCE, LOW INTEREST |
| FULLY INFORMED | Fondazione CHANGES | Conferenza dei Rettori delle Università italiane (CRUI) | CNR | |
| | | | Universities affiliated with Spoke 4 | |
| PARTIALLY INFORMED | Fondazione Scuola BB.CC. | | | |
| NOT INFORMED | | | | Consorzio Matera HUB |
| | | | | UnaEuropa |

**Figure 1.** The stakeholders identified for each dimension of the quadruple helix model, and clustered based on their relevance to the cultural heritage sector, their alignment with the project's strategic goals, and their degree of influence and involvement.

Despite the large number of stakeholders identified, the mapping activity conducted for engagement included clustering and, as a result, led to the elimination of some stakeholders. Discussions with workshop participants highlighted the need to focus on the most strategic stakeholders who were not already part of Spoke 4 activities. This approach ensured that the concluding scheme maintains strategic coherence with the other segments of the quadruple helix analysed.

Stakeholders, thus, were identified and clustered based on their relevance to the cultural heritage sector and their alignment with the project's strategic goals. Each stakeholder was then positioned according to its degree of influence and involvement, as shown in Figure 1, allowing for the development of targeted engagement strategies. The resulting framework supports continuous monitoring of participation through dedicated KPIs and guides optimisation of collaboration, dissemination, and the sustainability of Spoke 4 research outputs. This classification supported the definition of tailored engagement strategies and highlighted the most effective communication approaches for each stakeholder segment, ensuring alignment with their specific roles, interests, and the overarching goals of the CHANGES project. Starting from the output guidelines defined through the workshop, case studies leaders determined the stakeholders to be involved and the engagement strategies to be adopted for the implementation of their exploitation plans.

# Case studies

This section provides a brief overview of the case studies designed and conducted within Spoke 4. They cover cultural heritage institutions across Italy and fall into three categories, detailed in the following subsections: a *pilot case study*, used to develop acquisition and digitisation guidelines; *core case studies*, in which additional institutions and companies were directly involved in implementing the research activities; and *research case studies* conducted by Spoke 4 researchers. The guidelines developed through the pilot case study informed both the core and research case studies.

## Pilot case study

During the first months of the project, we identified a pilot case study to support the development of guidelines for Spoke 4 researchers involved in research that includes the acquisition processes of cultural heritage materials. This pilot case study could serve as a *sandbox* for involving a multidisciplinary group of experts to define approaches and methods for the acquisition, processing, optimisation, metadata inclusion, and online presentation of complex 3D assets. We selected a temporary exhibition held between December 2022 and May 2023 in the spaces of Palazzo Poggi at the University of Bologna, dedicated to Ulisse Aldrovandi (Balzani et al. 2024). Entitled *The Other Renaissance: Ulisse Aldrovandi and the Wonders of the World* (https://site.unibo.it/aldrovandi500/en/mostra-l-altro-rinascimento), the temporary exhibition provided a suitable setting to test the processes and workflows needed to enable visitors to enjoy a cultural event. The event has already concluded, with the meticulous 3D modelling of the exhibition’s six galleries, which contain more than 300 diverse cultural objects (e.g. natural history specimens, fossils, manuscripts, paintings, casts, and many other items), to be digitised using photogrammetry and structured-light scanning techniques. The overall goal was to recreate the exhibition experience digitally by developing a digital twin connected to the 3D and multimedia assets of the exhibited objects. We organised these resources into an online environment accessible through a range of devices, including computers, smartphones, tablets, and virtual reality headsets.

Such a temporary exhibition allowed us to work on several aspects of the acquisition and digitisation processes. These aspects included the impact that environmental, temporal, contextual and physical constraints (i.e. objects in glass cases, mandatory short time of acquisition, objects with peculiar materials such as fur and metal, etc.) have on the acquisition, and how we can structure the entire data flow of a complex cultural event such as an exhibition in a way that the digital data and metadata gathered are compliant with the FAIR principles. The ultimate goal was to define a workflow that allows us to preserve, in a digital environment, the physical relationships among the objects while maintaining and enriching the original curatorial narrative, supported by integrated audio guides. As a connected objective, we wanted the approach devised to establish a replicable case study for cultural heritage digitisation and valorisation practices, demonstrating how virtual and immersive experiences can effectively connect academic research with public engagement.

## Core case studies

Through the direct involvement of cultural heritage institutions interested in collaborating with Spoke 4, we have co-designed different lines of research, each mapped to one of the Types A-F presented in Section “Introduction”. The research planned for each case study pursued two complementary aims. On the one hand, we wanted to identify appropriate research scenarios to advance the application of digital and virtual technologies in the context of cultural heritage. On the other hand, we needed to align these research paths with the cultural heritage institution’s plans for the next few years, aiming for a mutually beneficial collaboration. These research lines were then published in two different Cascading Calls.

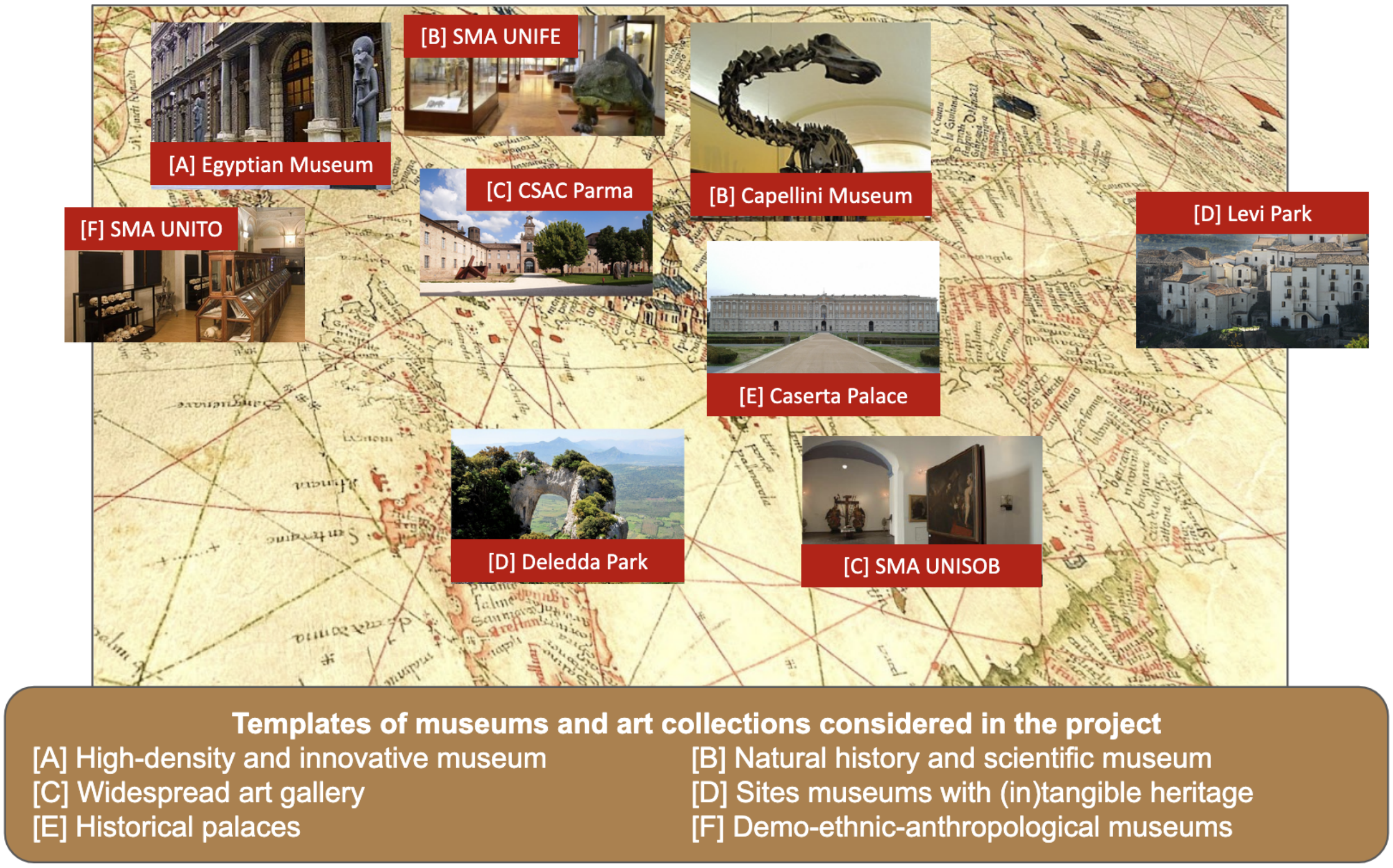

**Figure 2.** The nine core case studies conducted in Spoke 4 (with the related type specified), represented by the cultural heritage institutions we have collaborated with.

More than €2.5 million was allocated to implementing the research lines across nine *core case studies*, supplemented by co-funding from the participating organisations. The research involved 15 companies and 4 universities external to Project CHANGES, distributed across the entire country. Figure 2 summarises these case studies, listing the cultural heritage institutions involved and the related museum type. They are briefly described as follows.

**Type: *A*; CH institution: *Egyptian Museum in Turin*; Project title: *MEI: immersive Egypt project. Testing positive immersive virtual reality experiences to showcase the history and evolution of Egypt's historic and modern landscapes and their connection to the artefacts in the museum's collection.*** The *Museo Egizio Interattivo* (MEI) platform developed for the project supports curators in creating and delivering interactive stories inspired by ancient Egypt and grounded in scholarly research. The project aimed to achieve two main goals: ensuring historical and scientific accuracy in digital environments and characters by anchoring them in museum records, and offering visitors an immersive, interactive experience that enhances the museum's cultural heritage and introduces innovative ways to engage with it. The project developed software tools for both story production and managing interactive storytelling experiences within the museum context. Through a dedicated story editor, experts can now design and refine multilinear narratives in which audiences influence characters' decisions and shape the final outcome in an immersive environment (Mensa et al., 2025, 2026). Once defined, these stories can be adapted into playable formats suited to each venue's specific technological setup.

**Type:** ***B*****; CH institution:** ***Sistema Museale d'Ateneo Università di Ferrara (SMA UNIFE) – Museo di Paleontologia e Preistoria "Piero Leonardi" in Ferrara*****; Project title:** ***CHAMELEON: Open. Connected. Accessible | Digitisation and enhancement of museum environments and collections through enriched three-dimensional models*****.** CHAMELEON project wanted to develop digital tools to document, enhance, and share museum heritage, focusing on palaeontological collections in the *Piero Leonardi Museum* in Ferrara, which has pursued digitisation since 2013, mainly on smaller or non-displayed objects. The project expanded this effort by digitising over 30 significant specimens using 3D scanning and photogrammetry, creating semantically enriched models that combine scientific data with accessible information following Open Data standards. The developed platform aims to improve accessibility, both onsite and online, while enabling curators to update content and design flexible exhibitions. The system includes a back-end for managing content, AR/VR applications, and a scalable cloud infrastructure, which supports dynamic storytelling, enhances engagement, and allows new interpretative approaches to cultural heritage.

**Type:** ***B*****; CH institution:** ***Sistema Museale d'Ateneo Università di Bologna – Collezione di geologia "Museo Giovanni Capellini" in Bologna*****; Project title:** ***PALEOTWIN: virtual technologies at the service of the "Giovanni Capellini Museum" geology collection*****.** The *Museo Giovanni Capellini*, founded in 1881 with collections dating back to the 16th century, is housed in a historic building renovated in 1960. While preserving original 19th-century cabinets, this renovation reduced space, making exhibits crowded and harder for modern audiences to understand. The project wanted to introduce a multi-level technological system: management software to control diverse hardware outputs, digital avatars of past museum directors explaining their visions, immersive rooms for customizable digital exhibitions, and an augmented reality path through the historic displays. While coordinating multiple stakeholders was challenging, collaboration across diverse backgrounds enriched staff expertise and problem-solving approaches. The project created scalable digital exhibition systems adaptable to different museums, while engaging broader audiences through immersive, multisensory experiences that renew interest and reinterpret heritage.

**Type:** ***C*****; CH institution:** ***Centro Studi e Archivio della Comunicazione (CSAC) in Parma*****; Project title: PHYDOL-FHC: PHYGITAL CSAC.** Founded in 1968, the *CSAC* of the University of Parma is a major archive and museum holding around 12 million items across art, photography, design, and media. Since 2015, it has functioned as a multifunctional space combining archives, museum, and research. The project wanted to introduce advanced technologies such as 2D/3D scanning, photogrammetry, and virtual modelling to create *phygital* exhibition paths, combining tactile experiences for visually impaired users with immersive VR tools. The project also developed computational tools to transform artworks, particularly paintings, into three-dimensional models for tactile exploration, addressing the challenges involved in reconstructing 3D forms from 2D images. In addition, it improved linguistic accessibility by adapting museum texts to different levels of complexity for diverse audiences.

**Type:** ***C*****; CH institution:** ***Ente morale Istituto Suor Orsola Benincasa in Naples (SMA UNISOB)*****; Project title: MUSAD: a scalable and flexible digital platform for the valorisation of "widespread museums". The case of the museum system of the Istituto Suor Orsola Benincasa charity.** The project wanted to develop a scalable digital platform to enhance *widespread museality*. The project addressed challenges faced by smaller or peripheral institutions, such as limited accessibility and resources, by offering flexible tools for digitisation and virtual engagement that enable improved accessibility, sustainability, and audience engagement. The project developed three main components: a digital library for cataloguing and preserving artworks, an immersive platform that allows curators to create virtual exhibitions without technical skills, and an AI assistant that supports curators in designing exhibitions and narratives. Around 100 artworks were digitised in high resolution to demonstrate the system's potential, using advanced technologies such as 3D scanning and BIM modelling to ensure accurate digital replicas of exhibition spaces.

**Type:** ***D*****; CH institution:** ***Grazia Deledda Literary Park in Galtellì*****; Project title:** ***PALEGRADEVAL: the «Grazia Deledda» literary park: a valorisation project using virtual technologies*****.** The Grazia Deledda Literary Park in Galtellì, inspired by the Nobel Prize-winning writer, initially lacked technological and museum resources, offering only literary plaques despite its rich cultural value. Visitors had limited access to key locations, some of which were physically unreachable. The project introduced digitisation through high-resolution images, 360° content, and 3D models using innovative acquisition techniques and interactive environments integrated into maps, virtual tours, and mobile applications. A CMS-based system, databases, and touchscreen totems support content management and access across devices, and the team addressed key challenges, such as physical accessibility and the digital divide, through drones and intuitive interfaces. The system enhanced visibility, accessibility, and engagement, enabling remote exploration of cultural sites while supporting local stakeholders, promoting tourism, and positioning the park as an innovative cultural destination. The resulting platform integrated virtual tours, digital exhibition galleries, multimedia archives, and operational management tools into a coherent environment that supports both visitors and content managers.

**Type:** ***D*****; CH institution:** ***Carlo Levi Literary Park in Aliano*****; Project title:** ***VITALE: the «Carlo Levi» Literary Park: a valorisation project through virtual technologies*****.** The Carlo Levi Park in Aliano, inspired by the writer's exile and work, offers rich cultural assets but faced limited accessibility and a lack of digital resources. The project introduced extensive digitisation (photos, videos, 360° images, 3D models) to create virtual tours and interactive platforms. The project adopted technologies that enable immersive 3D environments and made them accessible via touchscreens and mobile devices. Additional tools experimented with within the project include web mapping, AI-based artwork recognition, photogrammetry, and audiovisual content like interviews and audio guides, with the goal of enhancing visibility, accessibility, and engagement with the Park and its content, allowing users to explore cultural sites remotely, offering a scalable model for other cultural contexts, benefiting local communities and institutions. Following validation and testing, the developed solutions were deployed and made operational within the park, supporting visitor experiences and cultural dissemination activities.

**Type:** ***E***; ***CH institution: Caserta Palace in Caserta***; **Project title:** ***RC/3DT: the queen's apartments in Luigi Vanvitelli's original design for the Royal Palace of Caserta.*** The project focused on creating a 3D digital twin of the Reggia di Caserta to support documentation, preservation, and interpretation of its architectural heritage. Targeting areas such as the Queen's Apartments and the third floor, whose original design has been altered over time, the project used advanced laser scanning and surveying techniques to produce accurate 3D models and 2D plans. These outputs contribute to historical research, restoration planning, and educational activities. A centralised digital repository has been developed to store and manage all data, ensuring accessibility and integration with existing museum systems. The project also experimented with augmented reality to enrich visitor experiences through additional narratives, strengthening the cultural value of this UNESCO site.

**Type:** ***F***; ***CH institution: Sistema Museale d'Ateneo Università di Torino (SMA UNITO) in Turin***; **Project title:** ***TAZEBAO: open source environment for the digital valorisation of museum collections on display, in storage and in archives. Experimentation with three-dimensional exhibition experiences, immersive virtual reality and interactive web.*** The project wanted to develop a scalable digital framework to document, enhance, and narrate the heritage of the University of Turin's Museum System. It focused on collections from the Lombroso Museum and MAET – such as Art Brut works, tattooed inmates' posters, and the Taíno cotton cemí – aiming to improve access to both visible and hidden heritage. The project introduced digitisation and interactive tools, including 3D exploration stations, VR installations, and a digital repository that enables personalised and multidimensional storytelling pathways, to support both research and public engagement, allowing users to explore complex cultural narratives across museums and supporting accessibility and digital sustainability, thus strengthening the role of museums as inclusive, innovative spaces for cultural exchange and education.

## Research case studies

The work conducted within the *core case studies* was complemented by additional research exploring the adoption of virtual and digital technologies in other cultural heritage institutions. We mapped these studies to the Types A–F introduced above. However, in contrast to the core case studies, we did not provide additional support for these additional research case studies from Cascade Calls. In practice, the Spoke 4 researchers involved in the study handled the research from design to implementation, while complying with all the guidelines and indications we set up for Spoke 4.

To gather these projects, we opened a specific internal call to Spoke 4 during the first year of Project CHANGES, tracking the additional efforts needed to address them and maintaining direct contact with their main proponents. This tracking activity generated a lot of interest and resulted in the proposal and implementation of an additional ten research case studies, listed as follows according to their type – Types B-F, except Type A, which did not receive any proposal for a research case study:

- Type B:

- Palazzo Poggi museum, such as the institute of sciences: the new permanent exhibition featuring virtual, multimedia, and historical elements – conducted by University of Bologna (UNIBO) researchers.
- Type C:
    - 3D technologies for early-modern sculpture and digital strategies for museums. Terracotta sculptures by Antonio Begarelli in the Galleria Estense, Modena – conducted by UNIBO researchers.
- Type D:
    - THERA – Theran epigraphic rubbings archive – conducted by Distretto Tecnologico per i Beni e le Attività Culturali (DTC) Lazio researchers;
    - The hidden legacy. Byzantine seals of the exarchal age in Italian museum collections – conducted by UNIBO researchers;
    - Digital strategies for enhancing cultural heritage: the Villa del Casale of Piazza Armerina, from the late antique building site to the museum collection – conducted by UNIBO researchers;
    - Museo storico italiano della guerra (Rovereto) – conducted by UNIBO researchers;
    - The castle of Gaeta: virtual technologies and new public history languages for the storytelling and musealization of memory – conducted by DTC Lazio researchers.
- Type E:
    - SPAFE – sculpture, painting and architecture fruition experience – conducted by DTC Lazio researchers.
- Type F:
    - Virtual environments as open archives to explore the cultural biographies of ethnographic artefacts at the Museo delle civiltà - MOCIV (Rome) – conducted by UNIBO researchers;
    - VIMMOB - virtual museum movable book (Panizzi Library) - mobile books digitisation and knowledge sharing for cultural accessibility – conducted by DTC Lazio researchers.

# Results

This section presents the main results obtained within Spoke 4 over its three-year duration. Results are organised around three distinct dimensions, which are detailed separately: (a) the design of workflows and implementation of tools and prototypes to support Spoke 4 research, in particular in the context of the case studies; (b) the meta-analysis of the results obtained by implementing the various case studies; (c) the quantitative and qualitative assessment conducted on usability, engagement and dissemination within the project.

## Workflows, tools and prototypes

One of the main objectives of Spoke 4 was to produce a set of workflows, tools, and prototypes for use in implementing the various case studies. However, because the research in these studies is heterogeneous, we mainly worked to address three distinct research dimensions, as summarised in Figure 3.

**Digitisation and digital heritage preservation.** Because digitisation is a prerequisite for all advanced interactions, the main goal was to move beyond traditional storage mechanisms, where DCH objects are available in specific locations without regulated public access, to implement and deliver interactive digital libraries and open repositories (SMA UNIFE, SMA UNITO, SMA UNISOB). In addition, it was central to experiment with new forms of interaction to maximise accessibility and inclusion, such as tactile 3D prints for the visually impaired (CSAC Parma) and Web-based exploration dynamics (Deledda and Levi Parks), and to enable the digital recovery of fragile environments or spaces that no longer exist in their original form (Reggia di Caserta), thus preserving what it was lost.

**Extended reality and immersive technologies.** Virtual reality systems were proposed for reconstruction to transport visitors to lost environments or inaccessible pasts (Reggia di Caserta, Museo Egizio di Torino). Instead, augmented reality approaches were used to enhance the experience by overlaying digital storytelling onto physical artefacts and enriching the on-site experience (CSAC Parma, SMA UNIFE, Museo Cappellini). Particular mention should be made of the *phygital* experiences implemented through seamless ecosystems in which digital content supports physical visits, shifting from static observation to dynamic interaction. In this context, engagement is increased by turning passive historical data into active, sensory-based narratives. Among the technologies adopted to enable these dynamics, ATON (https://osiris.itabc.cnr.it/aton/) (Fanini et al. 2021) was the primary open-source framework used in most case studies to create Web3D/WebXR applications that interact with cultural heritage objects and 3D scenes on the Web.

**Interactive storytelling and AI-supported narratives.** Storytelling was used to contextualise tangible and intangible cultural heritage artefacts through emotional journeys, historical anecdotes, and character-driven plots, also enabling digital access to museums' collections through digital storytelling paths (Egyptian Museum, SMA UNIFE). In addition, specific tools have been developed to enable curators and domain experts to annotate digitised museum assets and, then, facilitate the creation of interactive narratives by curators with an LLM-based generation tool, narratives that are then transformed into a playable story, coordinating assets, textual descriptions, and visitor feedback to enable interactive storytelling (Egyptian Museum).

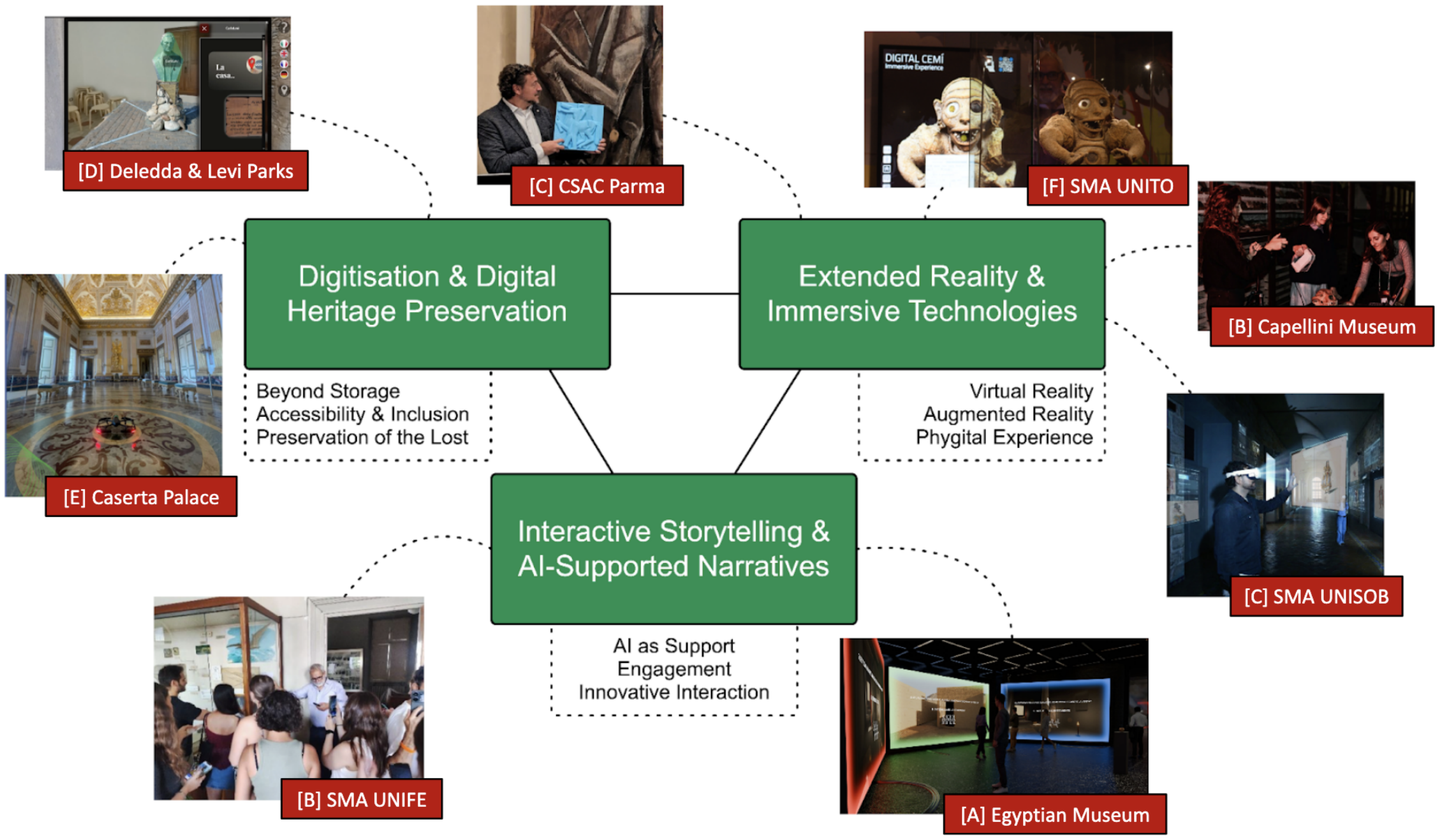

**Figure 3.** Three distinct levels of research conducted in the core case studies of Spoke 4, with some links between them according to the specificity of the work addressed.

## Data management and documentation

To ensure that this vast body of data and prototypes does not remain an isolated experiment but becomes a permanent scientific resource capable of turning digital valorisation into a widespread practice, we developed and used rigorous management and documentation methodologies. In particular, since the beginning of the project (and updated wherein) we defined a Data Management Plan (DMP) (Cavallo Perin et al. 2025) that has been considered since the beginning a strategic, dynamic, and collaborative framework to describe the categories of data and software produced within the project, specifying how they should be managed transparently in accordance with the FAIR principles (Wilkinson et al. 2016) and to the Recommendations on Open Science by UNESCO (2021), which put transparency as one of the necessary conditions for making digital heritage, and more generally any research object, a public, democratic, and lasting resource. Indeed, without transparent, well-structured data, the 3D models and prototypes we create today risk becoming unusable in just a few years, since only when data is open, accessible, and understandable can it move beyond laboratories and truly be shared with the wider community (Barzaghi et al. 2025b).

To pursue these goals in the context of Spoke 4 research, we needed to bring together a highly diverse range of expertise, given the complexity and multidisciplinary nature of cultural heritage research today. Thus, we included in our research team:

- *3D digitisers*, to enable the acquisition of data using different technologies, process it, and produce various versions of 3D models using specific software, depending on needs related to conservation, research, or public access;
- *metadata experts*, bringing in a crucial contribution to enable 3D models not to be invisible to the community, by ensuring the availability of structured, machine-readable metadata that enables discovery, citation, and scientific traceability of the data, both for the original physical object and its digital representations;
- *museum curators*, to ensure the scientific context by validating interpretation, the relationship with the physical object, and the cultural integrity of the data;
- *legal experts*, to address the complex issues of 3D authorship and open licensing;
- *Web infrastructure specialists*, including those who develop/manage web-based applications for integrating 3D assets and environments, and those who design systems that make these models semantically queryable in dialogue with the produced metadata.

While when we think about the concept of *infrastructure* we often implicitly refer to the hardware and tools needed to address specific research activities, the *humans* who use such hardware and tools are the keystone of such an infrastructure and, in the context of the work conducted in Spoke 4, none of the aforementioned skills is secondary to others. All these expertises are fundamental to successfully implementing infrastructures devoted to the valorisation and preservation of cultural heritage with virtual and digital technologies. Indeed, implementing Open Science principles for 3D cultural artefacts works only as an integrated process, where each professional figure is an essential piece in making data truly FAIR-compliant through the creation of a shared workflow (Barzaghi et al. 2024a, 2025a).

## Technical standpoint

From a technical point of view, when we move from the physical object to the digital one, it is important to emphasise that a 3D model is never a simple photostatic 1:1 reproduction of the original physical object, but rather a true creative artefact. Indeed, a 3D model is complex data, resulting from technical choices, different acquisition methodologies (which also depend on the technique, e.g. either scanner or photogrammetry, and the conditions and intrinsic characteristics of the object to be acquired), data-cleaning algorithms, and scientific interpretation. For this reason, a methodological gap always exists between the real Cultural Heritage Object (CHO) and its digital representation. To manage this gap scientifically, preserving different derived versions (each with a specific function) makes the process of authorial intervention during modelling more transparent, allowing direct comparison between versions and tracking the evolution of the work:

- we started from the *raw material* (RAW), which refers to the pure acquisition data;
- we moved to a *processed version of the raw* material (RAWp), which is the preliminary result of an initial processing performed via appropriate software without interpolation or geometric correction;
- we arrive at the refined *digital cultural heritage object* (DCHO), i.e. the refined digital object where geometric and/or texture issues are resolved;
- finally, we produce an *optimised version of the DCHO* (DCHOo) for a specific purpose – in our case, its real-time web visualisation.

Beyond these versions, additional creative variations may be created, where digital simulations and restorations can be built from the 3D acquired from a given physical object. For these reasons, the digital product constitutes a new intellectual entity and belongs to the *professionals who created it*, although it remains inseparable from the original physical cultural object.

Each 3D object must also be accompanied by metadata documenting the creative and technical processes involved, ensuring that future users can understand its origin, accuracy, and context. In Spoke 4, we orchestrated a dialogue between two distinct biographies: while the metadata of the physical object reconstructs its historical biography, the digital data documents its technical biography and the digitisation process. To give concrete form to this, we created and adopted specific guidelines and technologies (Barzaghi et al., 2024b, 2026) that uses the Semantic Web paradigm and recognised international standards such as CIDOC-CRM (Doerr, 2003) for creating a publishing machine-actionable metadata ready to be rendered and queried by semantic platforms.

The entire methodological framework presented here (Bordignon et al., 2024) was derived from our pilot case study, i.e. the creation of the digital twin of the Ulisse Aldrovandi temporary exhibition (Balzani et al., 2024). Indeed, the pilot served as a shared ground to consolidate a common pipeline and define best practices for the digitisation of cultural heritage, which has been used in Spoke 4 when similar contexts had to be handled in the other case studies. The main objective was to generate 3D models that maintain visual credibility while undergoing geometric optimisation for use in a Web3D environment on ATON, without sacrificing data reliability and authenticity.

## Meta-analysis of the case studies

The meta-analysis results focus on specific analytical dimensions, including museum typology, technological configurations, organisational adoption, and the role of core and research case studies. Table 3 summarises the results of such a cross-case analysis, namely the patterns, convergences and divergences discussed above, making explicit the relationships between museum typologies, technological roles and implementation strategies. Beyond synthesising patterns and divergences, the cross-case comparative analysis also provides the basis for identifying the main technological gaps and for outlining future lines of research grounded in the empirical evidence of the core case studies. The following subsections discuss these dimensions in more detail.

| **Type** | **Case study** | **Technological roles** | **Technological configuration** | **Familiarisation model** | **Organisational embedding** | **Experimentation status** | **Key challenges** |
|---|---|---|---|---|---|---|---|
| A | C | Immersive storytelling | Annotation editor, story editor, stage manager | Co-creation workshops | Medium-High | Structured user testing in temporary setting analysis ongoing | System integration, infrastructure |
| B | C+R | Scientific | 3D digitisation | Research-based | Medium | Short-term post- | Data volume, |
| | | documentation | and visualisation pipelines | training | | opening observation; partial evaluation | legacy systems |
| C | C | Curatorial infrastructure | Archive, virtual exhibitions, AI assistant | Hands-on workshops + UX feedback | High | Staff-centred structured evaluation completed; public testing partially addressed | Metadata quality, governance |
| C | R | Analytical research | Advanced 3D analysis | Academic training | Low-Medium | No systematic user-facing experimentation | Transfer to deployment |
| D | C | Territorial access | Lightweight 3D (including innovative 3D modelling techniques), maps, responsive web applications | Manuals and demos | Medium-High | In situ experimentation during inaugurations and use | Performance, legal issues |
| E | C | Management and conservation | Digital twin, GIS, AR applications | Replicated training | High | Dual evaluation: staff UX assessment and public demos | Coordination, governance |
| F | C | Semantic repository | Open repository and modular interfaces | Progressive familiarisation | Medium | Extended visitor experimentation over exhibition period | Ethical and legal constraints |
| F | R | Exploratory research | Experimental prototypes | Research-based training | Low-Medium | Experimentation deferred (exhibitions pending) | Institutional timing |

**Table 3.** Comparative synthesis across museum Types A-F, considering both core case studies (C) and research case studies (R).

## Museum types, technologies and their configurations

The cross-case analysis conducted confirms a strong relationship between museum typology and the technological configurations adopted across both core and research case studies. While similar technological families recur across cases (i.e., 3D digitisation, immersive environments, Web-based platforms, and AI-assisted tools), their role, complexity, and intended function vary consistently with institutional context, spatial constraints, and organisational needs. Overall, the results show that museum typology shapes the form of technological solutions and their conceptual role within institutional practices, as briefly presented in the following paragraphs.

In Type A museums, technologies are primarily conceived as *immersive narrative infrastructures*. Digital solutions take the form of integrated production pipelines enabling the authoring, orchestration, and delivery of complex interactive experiences for large and diverse audiences. The emphasis is placed on content programmability, narrative branching, and real-time orchestration.

Instead, technologies oriented toward *scientific documentation, interpretation, and mediated access to complex collections* characterise Type B museums, while immersive and narrative components play a limited role compared to Type A. The focus here is more on data reliability and integration with scientific workflows, advanced digitisation pipelines, data-driven visualisation, and interactive systems that support research activities and public understanding. At the same time, user-facing interfaces are often shaped by infrastructural constraints and by the need to balance scientific rigour with accessibility.

In Type C art galleries, the technologies are configured as *curatorial and infrastructural ecosystems*, to support distributed collections, multiple curators, and heterogeneous exhibition contexts. Often, such systems combine digital archives, virtual exhibition tools and semantic or AI-assisted components, with strong attention to interoperability standards, metadata quality, modularity, and open-source solutions to ensure scalability and long-term sustainability.

*Territorial access and interpretation systems* define the technological configurations adopted in Type D scenarios, favouring lightweight 3D reconstruction pipelines, immersive tours linked to geographic routes, and delivery through web-based or mobile solutions that support *in situ* exploration of widespread cultural landscapes. Specific performance constraints, accessibility, and ease of maintenance strongly shape the technological choices, leading to solutions that minimise hardware dependencies and support deployment across dispersed points of interest.

Type E locations often feature *digital twins and professional-grade management platforms*. In these contexts, technologies focus on documentation, conservation, monitoring, and decision support. Web platforms integrating point clouds, GIS layers, and architectural data are complemented by mobile or AR applications designed to support on-site experience.

Finally, Type F museums adopt *open, reconfigurable infrastructures* that combine semantic repositories, modular interfaces, and interactive installations. These cases are distinguished by the need to manage heterogeneous collections and sensitive materials, which introduces additional ethical, legal, and governance constraints and places particular emphasis on flexibility, access control, and editorial oversight.

## Familiarisation with technologies

In the case studies analysed, the familiarisation strategies used to introduce these technologies to museum staff and other stakeholders are not uniform; instead, they reflect the expected degree of staff autonomy, the complexity of the technological systems, and the institutional role of the digital tools. Importantly, training and familiarisation emerge not as ancillary activities but as critical enablers of implementation sustainability and effectiveness. In particular, the introduction of such technologies follows distinct models which are closely aligned with both museum type and the nature of the implemented systems.

A first recurring model is *co-creation familiarisation*, typically adopted when technologies support either creative or curatorial authoring processes. In these cases, training is embedded within workshops and collaborative activities involving museum professionals, developers and designers. Familiarisation focuses less on procedural instruction and more on enabling participants to actively shape content, narratives and workflows. This approach is prevalent in Types A, B, C, and F, where technologies function as tools for storytelling or exhibition design and where iterative feedback is integral to system refinement.

A second model is *operational enablement*, which is common in territory-based or platform-oriented deployments. In this context, familiarisation is structured around manuals, tutorials, demonstrations, and hands-on sessions aimed at ensuring that local staff and external stakeholders can autonomously operate, update and maintain the systems. This model is particularly evident in Type D, where municipal actors, guides or local administrations are expected to manage digital tools in the long term with limited technical support. In such cases, operational enablement is achieved through targeted online (and in-person) training sessions with administrations and cultural organisations to demonstrate both the graphic design of the web applications and how to operate them. These activities are supported by concise operating manuals to ensure autonomy in managing content updates, and by simple applications for monitoring system operation.

A third strategy, observed especially in Type E, is *scaled and replicated training*. In these cases, initial training is delivered to a core group of expert users and subsequently replicated for a broader range of institutional staff. This approach reflects the professional and operational nature of the technologies involved and the need to embed them within established organisational structures. Training effectiveness is sometimes verified through questionnaires or structured feedback, indicating a more formalised adoption process.

## Maturity of case studies

The distinction between core case studies and research case studies proves analytically relevant when interpreting implementation outcomes. Core case studies generally present technologies that are already deployed, accompanied by explicit strategies for training, organisational adoption, and maintenance. In these cases, the focus is on implementation readiness, institutional embedding, and, where possible, early evaluation. Instead, research case studies are more often characterised by experimental or exploratory configurations, often depending on data availability, exhibition schedules, or broader institutional constraints. While they adopt similar technological approaches to core case studies, their primary contribution lies in methodological innovation, proof-of-concept development and scholarly investigation, rather than immediate service deployment.

This distinction suggests that core and research case studies should not be evaluated by the same criteria. Instead, they represent different positions along an implementation continuum: research case studies inform future development and refinement, while core case studies test adoption, governance, and sustainability in real operational contexts.

## Experimentation and evaluation

By the end of the project (April 2026), experimentation with end users cannot be systematically compared across all the case studies. In particular, the Type A core case study held in conjunction with the Egyptian Museum has experienced delays in completing its permanent immersive infrastructure, resulting in evaluation activities being conducted in temporary settings. Similarly, other case studies report experimentation as planned or ongoing, often contingent on either exhibition readiness or institutional reopening. Moreover, experimentation sessions differ substantially in terms of timing, target users, evaluation instruments and maturity of results, which limits direct cross-case comparability at the level of outcomes.

In the research case studies, experimentation is often deferred or limited to methodological validation because of pending exhibitions, ethical constraints, or the exploratory nature of the research. Instead, in the core case studies, experimentation is primarily formative and validation-oriented. High-density museum contexts (Type A) conducted structured user testing with real visitors and staff using standardised UX instruments, producing evidence of high engagement and perceived quality, although testing took place in a temporary setting (Antonino et al., 2026). Natural history and scientific museums (Type B) report short-term post-opening observation and system log analysis, highlighting positive interaction with selected components alongside usability and technical issues. In widespread art gallery contexts (Type C), experimentation is mainly staff-centred and highly structured, combining qualitative feedback with standardised UX questionnaires to assess curatorial platforms, while public-facing evaluation remains partial. Site museums and landscape contexts (Type D) rely on in situ experimentation with visitors, using structured questionnaires and qualitative feedback reporting high satisfaction and recommendation intent. Historical palaces and UNESCO sites (Type E) adopt a dual evaluation approach, combining staff-oriented UX assessment of management platforms with large-scale public demonstrations of experimental applications, both yielding positive results. Demo-ethno-anthropological museums (Type F) present extensive experimentation, with large visitor samples, combined UX instruments and web analytics collected over extended exhibition periods, resulting in high engagement and positive reception.

Overall, experimentation results across Types A-F show a predominantly positive orientation, particularly in terms of engagement, perceived value and likelihood of recommendation. Nevertheless, the heterogeneity of evaluation designs and implementation stages confirms that experimentation outcomes should be interpreted as indicators of implementation maturity and learning processes rather than as directly comparable performance metrics.

## Emerging issues

The cross-cutting issues identified through the comparative analysis can be interpreted as technological and organisational gaps emerging from the implementation of the core case studies. Rather than isolated shortcomings in specific solutions, these gaps reflect structural conditions that affect scalability, sustainability, and long-term impact across different museum typologies.

These conditions include challenges related to long-term sustainability, such as dependence on metadata quality, availability of human experts in the technological infrastructure and resources provided, governance of digital infrastructures, and ethical or legal constraints affecting data publication and reuse. Their recurrence across museum types and case categories points to shared areas where further research and strategic development are needed.

## Usability assessment, user engagement, dissemination

A configurable web dashboard – capable of tracking and accounting activities, and filtering them by case study and partner – has been developed to support the evaluation and tracking of:

- exploitation – by tracking which stakeholders are involved, by whom and how, enabling aggregated counts;
- dissemination – by recording the types of outputs, links, and responsible partners, enabling quantitative monitoring;
- user assessment – by collecting and tracking evaluation activities.

Spoke 4 effectively engaged a wide and diversified community of stakeholders, established a robust workflow for the digital dissemination and enhancement of cultural heritage, and ensured a high level of scientific production and outreach.

The exploitation activities led to outstanding results in scientific dissemination, with 51 publications and participation in 12 national and international conferences within the project timeframe, largely surpassing the initially defined targets. At the same time, the communication strategy ensured continuous visibility and engagement through a sustained presence on digital channels, newsletters and institutional platforms. Overall, the results confirm the effectiveness of the strategic framework adopted by Spoke 4, exceeding initial expectations beyond the project targets in both quantitative and qualitative terms.

The workshop “CultureInDialogue” was a key milestone in Spoke 4's stakeholder engagement strategy, directly supporting the project’s objective of engaging relevant communities and stakeholders in developing and disseminating its results. By bringing together 27 representatives from the main Spoke 4’s research units and core case studies, from 3 of the 4 classes of the quadruple helix (Academia, Government, and Industry/Professionals), the four-hour collaborative session strengthened the stakeholder mapping process and established a shared framework for future involvement and dissemination activities. The participatory discussion refined the stakeholder mapping across all quadruple helix sectors, identifying 19 stakeholders in Government, 20 in Academia, 19 in Civil Society, and 11 in Industry.

Regarding the identification of a professional network, the target has been fully achieved through the establishment of a structured, continuous, and strategic collaboration with TICHE Foundation - National Technological Cluster for Cultural Heritage Technologies (https://www.fondazionetiche.it/). Within this framework, a significant B2B engagement activity was carried out, fostering direct interaction between case-core managers and potentially interested companies. Furthermore, Spoke 4 is naturally integrated and aligned with Foundation

CHANGES, which can be identified as an additional strategic professional network, further strengthening the overall ecosystem of collaboration and innovation.

| KPI | Description | Result |
|---|---|---|
| User Involvement Volume | Number of users involved in the user testing activities | 550 |
| Use Case Evaluation Coverage | Percentage of core use cases for which user testing was conducted | 60% |
| Average Usability Assessment | Average evaluation of perceived usability of users by means of SUS | 57,06 |
| Usability Assessment Coverage | Percentage of user testing activities in which usability was assessed, regardless of the specific instrument adopted | 50% |
| User Experience Quality Assessment | Average evaluation of overall user experience quality by means of UEQ-S | 1,26 |
| User Experience Quality Assessment Coverage | Percentage of user testing activities in which overall user experience quality was assessed | 40% |
| Endorsement | Average evaluation of users' predisposition to recommend the developed solutions as an early signal of potential adoption by means of NPS | 53,41 |
| Endorsement Coverage | Percentage of user testing activities in which predisposition to recommendation was assessed | 30% |

**Table 4.** User engagement and usability assessments KPIs and the related results.

The monitoring indicators associated with (a) the development of exploitation strategies for maximising reusability and enabling the scaling up of narratives and prototypes at the national level, and (b) the assessment of the impact of digital dissemination approaches on potential audiences, provide a consolidated picture of the extent to which structured user-centred evaluation practices were implemented across the project.

As summarised in Table 4, from a coverage perspective, structured user testing activities were conducted in 60% of the core case studies, indicating that most cultural heritage venues integrated formal qualitative validation processes. Usability assessment, irrespective of the specific instrument adopted, was implemented in 50% of user testing activities, while overall user experience quality was assessed in 40% of cases. We measured predisposition to recommendation (endorsement) in 30% of testing activities. As previously clarified, these percentages serve a descriptive monitoring function, capturing the extent and intensity of formalised user-centred evaluation effort across the project. In terms of user involvement volume, a total of 550 users participated in testing activities. This number reflects the cumulative participation recorded across the various evaluation contexts within the project timeframe.

Table 4 averages were computed as unweighted means across evaluation instances rather than weighted by the number of respondents. This methodological choice avoided the disproportionate influence of larger-scale case studies and preserved analytical balance across heterogeneous use cases with different sample sizes and validation contexts. The average SUS score of 57.06, computed across two evaluation instances (within two case studies) for 32 respondents, indicates that, for the considered case studies, usability requires further refinement. Indeed, the SUS score below the 68 threshold suggests that aspects such as interaction clarity, efficiency, and ease of use could be strengthened in subsequent iterations.

In contrast, the average UEQ-S score of 1.26, calculated across eight evaluation instances (within three case studies) with 330 respondents, reflects a clearly positive overall user experience for the evaluated use cases. Indeed, under established UEQ interpretation frameworks, values above 1 indicate distinctly favourable evaluations, suggesting users perceived the solutions as supportive and experientially satisfying across both pragmatic and hedonic dimensions. The average NPS of 53.41, computed across ten evaluation instances (within three case studies) with 524 respondents, reflects a positive predisposition to recommend the considered solutions. In NPS interpretation scales, values above 50 are generally considered excellent and indicate strong user advocacy.

It is also important to note that, when the recommended instruments (SUS, UEQ-S, NPS) were not adopted, alternative evaluation tools were used to address context-specific needs. In particular, some case studies, especially those developing tools intended to support professional actors rather than the general public, integrated additional constructs such as perceived usefulness. This dimension is particularly relevant in contexts involving museum professionals and digital curators, where the tool's added value in supporting expert workflows is a central evaluation criterion. The use of complementary instruments therefore reflects methodological adaptation to different user profiles and validation objectives, rather than a deviation from the evaluation framework.

# Discussions and conclusions

Within the research conducted in Spoke 4, we aimed to design an acquisition and digitisation workflow as closely aligned as possible with the principles of Open Science, as per funder's mandate, placing reproducibility and interoperability at its core. It should be emphasised that we do not propose our framework as a either universal or rigid solution. Indeed, we are aware that every organisation and institution may already have established workflows, and substituting or integrating them with a new one is far from straightforward. However, we believe our approach adds value where methodological transparency and collaboration between institutions become essential. In this way, the pilot case study on the Ulisse Aldrovandi temporary exhibition presents itself as a scalable and shareable model: a concrete attempt to make the digitisation process not just a technical operation, but an open, verifiable scientific pathway oriented toward knowledge sharing, transforming a temporary event into a permanent and reproducible digital legacy – that can be either downloaded and run locally (Ammirati et al., 2026) or accessed online at https://w3id.org/changes/4/aldrovandi/dt/.

The meta-analysis of the case studies enabled us to conduct a qualitative cross-case synthesis to understand how technological solutions are implemented, adopted, and embedded across heterogeneous museum contexts. We show that museum typology plays a decisive role in shaping not only technological choices but also organisational strategies, training models, and pathways to sustainability. In addition, the results show that similar technological families assume different roles depending on institutional context, confirming that a successful implementation depends less on the intrinsic sophistication of technologies than on their alignment with contextual constraints, institutional missions and organisational capacities.

By reading together the different data gathered from the case studies, it is possible to identify a set of technological and organisational gaps that go beyond single cases and reflect shared challenges across museum contexts. These gaps primarily concern long-term sustainability, metadata quality, capacity building, governance and infrastructural autonomy. Based on these identified gaps, several further lines of research emerge as critical for advancing digital cultural heritage practices. These include developing interoperable and modular solutions, explainable and controllable uses of artificial intelligence, longitudinal evaluation of user experience, and inclusive design approaches capable of supporting diverse institutional and social contexts.

The data gathered within the project timeframe documented the success of the engagement, exploitation, and dissemination strategy implemented within Spoke 4, demonstrating that the objectives initially set were not only fully achieved but, in many cases, significantly exceeded according to all the dimensions considered – i.e. active stakeholder engagement, the outcomes of scientific dissemination, and communication and outreach actions. This effort enabled the development of a solid, cohesive ecosystem recognised at both national and international levels, as confirmed by the prestigious CHANGES Awards – created by the CHANGES Foundation to recognise projects within Project CHANGES that successfully combine scientific excellence, innovation capacity, and potential impact on cultural heritage and society – received by Spoke 4 research groups in 2025 and 2026.

The activities described do not end with project completion. A key future direction concerns the transition from prototype-based experimentation to stable, long-term infrastructures. This implies investing in interoperable, modular platforms that adapt to diverse museum typologies while remaining independent of proprietary systems. In parallel, strengthening metadata quality and semantic standards will be essential to enable cross-institutional integration, data reuse, and the emergence of meta-museum ecosystems.

Future efforts should also focus on structured training models for museum professionals, alongside the definition of clear governance frameworks that address ethical, legal, and organisational aspects – particularly in relation to sensitive collections and the use of artificial intelligence. Longitudinal evaluation of user experience will also become increasingly central. Moving beyond pilot testing, future research should adopt continuous assessment methodologies to understand how different audiences engage with digital environments over time, with a strong emphasis on inclusive design and accessibility across social and cognitive diversity.

Together, Spoke 4's research has laid strong foundations to ensure the project's legacy remains a living, accessible, and strategic resource for the future of cultural institutions at national and international levels. Indeed, the consolidation of the ecosystem, built and supported by robust dissemination strategies, FAIR-compliant data management, and international recognition, opens the door to scaling these approaches beyond the project. For instance, the winning project of the PON RIC 2021/27, entitled HERITAS (https://www.fondazionechanges.org/2026/07/14/heritas-al-via-il-progetto-che-rafforza-lecosistema-dellinnovazione-per-il-patrimonio-culturale/), has among its explicitly stated objectives to bring the technologies we developed in Spoke 4 up to TRL 8 to transform research prototypes into pre-commercial and scalable solutions that can help consolidate Italy as a world leader in the ethical and inclusive application of advanced technologies for cultural heritage.

As a final remark, the primary goal reached by Spoke 4 probably goes beyond the detailed analysis we reported in this article. On the one hand, it was crucial to have produced such a long list of research objects, from workflows to 3D models of cultural heritage objects, from prototypes to instantiations of exhibitions using or mixing up digital and virtual technologies to enhance accessibility and valorisation of cultural heritage, to prove quantitatively and qualitatively the success of the project. On the other hand, the most significant result was that, thanks to the involvement of several institutions and companies distributed across the whole country, we have been able to create an unprecedented and decentralised network of researchers with a huge variety of skills applied to cultural heritage, out of an arid environment that, in the past, favoured primarily competitive and monodisciplinary behaviours rather than cooperative and multidisciplinary interchanges at scale. The creation of such a large, *human decentralised infrastructure* (HDI) was possible because of an exceptional, probably one-of-a-kind, allocation of funds for research in the Italian cultural heritage domain, and it is probably the biggest legacy of Spoke 4 and, more broadly, Project CHANGES. We, as researchers and institutions involved in this project, have the moral duty of not letting this opportunity slip away, forcing us to continue feeding such an environment we have built to advance cultural heritage research and return positive influence to the society.

# Acknowledgements

We want to extend our gratitude to all the companies and institutions that have been directly involved in the project via the Cascading Calls and that have brought their expertise to finalise the implementation of all the core case studies. These are (in alphabetic order): BASILINK ART SRLS, Digitarca SRL, ENERGICAMENTE SRL, FORUM ENGINEERING, GLOSSA SRL, IMAGO, INCEPTION SRL, MEDIASOFT SRL, NAIS SRL, No Real Interactive SRL, Politecnico di Torino, RIBES DIGILAB SRL, Risviel SRL, Robin Studio, Tecno Art SRL, TP SRL, Università degli Studi di Ferrara, Università degli Studi di Napoli L'Orientale, and Università degli Studi di Parma.

This article has been a multi-authored contribution that involved, as proper authors, all the scholars who have either contributed to any of the project deliverables and/or published some articles and data pertaining the research conducted within Spoke 4, which have been fundamental to reach all the results described in the previous sections. However, we are fully aware that there were more people – from the cultural heritage institutions involved and the CHANGES Foundation, and other researchers, students, administrators, IT experts, from all the entities (universities, research centres, and companies) involved in the project – who have worked behind the scenes to enable us the fully accomplishment of Spoke 4's objectives. Thus, with these words, we would like to thank them all for their invaluable contribution.

# Funding

This work was funded by Project PE 0000020 CHANGES - CUP B53C22003780006, NRP Mission 4 Component 2 Investment 1.3, Funded by the European Union - NextGenerationEU.

# Declaration of Use of Generative AI

No generative AI systems were used to generate or analyse data, or draft the scientific content of this manuscript. AI-assisted tools were used only for spell-checking and minor grammar suggestions on the English text.

# References

Amitrano, C. C., Montaldo, S., & Stalteri, C. (2025, October 1). *Guide to implementation* [Template]. Zenodo. https://doi.org/10.5281/zenodo.18608699

Ammirati, L., Barzaghi, S., Bonifazi, F., Bordignon, A., Casadei, V., Cipriani, L., Colitti, S., Collina, F., Daquino, M., Fabbri, F., Fanini, B., Fantini, F., Ferdani, D., Fiorini, G., Forte, A., Giacomini, F., Girelli, V. A., Gualandi, B., Heibi, I., … Travaglini, L. (2026, May 29). *Aldrovandi Digital Twin* (Version 1.0.0) [Digital Twin]. Zenodo. https://doi.org/10.5281/zenodo.20451916

Antonino, R. A. S., Damiano, R., Bottino, A., Fallucchi, F., Mensa, E., & Ferraris, E. (2026). Collective Spatial Interaction for Immersive Museum Storytelling: The MEI (Museo Egizio

Interattivo) Experience. *Proceedings of IEEE-CH Cyber Humanities 2026*. IEEE Cyber Humanities 2026 (IEEE-CH 2026), Venice, Italy.

Aprea, Diego, Bagnoli, Martina, Baldi, Francesco, Baldinotti, Stefania, Barbuto, Alessandra, Bellingeri, Luca, Benedetto, Stefano, Benintende, Angela, Birrozzi, Carlo, Boi, Valeria, Buttò, Simonetta, Casula, Vassili, Ciancio, Laura, Coco, Alessandro, Conticelli, Valentina, Corrao, Alfredo, Cundari, Chiara, De Chirico, Fabio, De Luca, Martina, … Zuchtriegel, Gabriel. (2023). *Piano Nazionale di Digitalizzazione del Patrimonio Culturale 2022-2023* (Moro, Laura, Ed.; Linee Guida Versione 1.1). Ministro della Cultura. https://digitallibrary.cultura.gov.it/wp-content/uploads/2023/04/PND_V1_1_2023_v2.pdf

Avviso n. 341 Del 15-03-2022, 341, Ministero dell'Università e della Ricerca (2022). https://www.mur.gov.it/it/atti-e-normativa/avviso-n-341-del-15-03-2022

Balzani, R., Barzaghi, S., Bitelli, G., Bonifazi, F., Bordignon, A., Cipriani, L., Colitti, S., Collina, F., Daquino, M., Fabbri, F., Fanini, B., Fantini, F., Ferdani, D., Fiorini, G., Formia, E., Forte, A., Giacomini, F., Girelli, V. A., Gualandi, B., … Vittuari, L. (2024). Saving temporary exhibitions in virtual environments: The Digital Renaissance of Ulisse Aldrovandi – Acquisition and digitisation of cultural heritage objects. *Digital Applications in Archaeology and Cultural Heritage*, *32*, e00309. https://doi.org/10.1016/j.daach.2023.e00309

Barzaghi, S., Bordignon, A., Collina, F., Fabbri, F., Fanini, B., Ferdani, D., Gualandi, B., Heibi, I., Mariniello, N., Massari, A., Massidda, M., Moretti, A., Peroni, S., Pescarin, S., Rega, M. F., Renda, G., & Sullini, M. (2025a). A reproducible workflow for the creation of digital twins in the cultural heritage domain. *Transformations: A DARIAH Journal*, *1*. https://doi.org/10.46298/transformations.14773

Barzaghi, S., Bordignon, A., Gualandi, B., Heibi, I., Massari, A., Moretti, A., Peroni, S., & Renda, G. (2024a). A Proposal for a FAIR Management of 3D Data in Cultural Heritage:

The Aldrovandi Digital Twin Case. *Data Intelligence*, *6*(4), 1190–1221. https://doi.org/10.3724/2096-7004.di.2024.0061

Barzaghi, S., Bordignon, A., Gualandi, B., & Peroni, S. (2025b). Enlightening the Black Box of Humanities Research. Methodological Documentation as a Way to Transparency and Accountability of Digital Exhibitions. *Umanistica Digitale*, *9*(20), 97–114. https://doi.org/10.6092/issn.2532-8816/21177

Barzaghi, S., Heibi, I., Moretti, A., & Peroni, S. (2024b). Developing Application Profiles for Enhancing Data and Workflows in Cultural Heritage Digitisation Processes. In G. Demartini, K. Hose, M. Acosta, M. Palmonari, G. Cheng, H. Skaf-Molli, N. Ferranti, D. Hernández, & A. Hogan (Eds), *The Semantic Web – ISWC 2024* (Vol. 15233, pp. 197–217). Springer. https://doi.org/10.1007/978-3-031-77847-6_11

Barzaghi, S., Moretti, A., Heibi, I., & Peroni, S. (2026). CHAD-KG: A Knowledge Graph for Representing Cultural Heritage Objects and Digitisation Paradata. *International Journal on Semantic Web and Information Systems*, *22*(1). https://doi.org/10.4018/ijswis.403910

Bordignon, A., Barzaghi, S., Collina, F., Fabbri, F., Fanini, B., Ferdani, D., Marinello, N., Moretti, A., Rega, M. F., & Sullini, M. (2024). *Guidelines for the digitisation of museum and art collections*. Zenodo. https://doi.org/10.5281/zenodo.14249936

Brooke, J. (1996). SUS: A 'Quick and Dirty' Usability Scale. In P. W. Jordan, B. Thomas, I. L. McClelland, & B. Weerdmeester (Eds), *Usability Evaluation In Industry* (pp. 207–212). CRC Press. https://doi.org/10.1201/9781498710411-35

Carayannis, E. G., & Campbell, D. F. J. (2009). 'Mode 3' and 'Quadruple Helix': Toward a 21st century fractal innovation ecosystem. *International Journal of Technology Management*, *46*(3/4), 201. https://doi.org/10.1504/ijtm.2009.023374

Cavallo Perin, R., Gualandi, B., Paruzzo, F., & Peroni, S. (2025). *Data Management Plan: Final version* (Deliverable D10 v1.0; Project CHANGES - Spoke 4). Alma Mater Studiorum - Università di Bologna. https://doi.org/10.5281/zenodo.14955816

Commission Recommendation of 27 October 2011 on the Digitisation and Online Accessibility of Cultural Material and Digital Preservation, 2011/711/EU, Official Journal of the European Union (2011). http://data.europa.eu/eli/reco/2011/711/oj

Council of the European Union. (2025). *Council Recommendation of 24 June 2025 on the European Research Area Policy Agenda 2025-2027 (Text with EEA relevance)* (Recommendation C/2025/3593). European Union. https://eur-lex.europa.eu/legal-content/EN/TXT/PDF/?uri=OJ:C_202503593

Davis, F. D. (1989). Perceived Usefulness, Perceived Ease of Use, and User Acceptance of Information Technology. *MIS Quarterly*, *13*(3), 319–340. https://doi.org/10.2307/249008

Davis, F. D., & Granić, A. (2024). *The Technology Acceptance Model: 30 Years of TAM*. Springer International Publishing. https://doi.org/10.1007/978-3-030-45274-2

Doerr, M. (2003). The CIDOC Conceptual Reference Module: An Ontological Approach to Semantic Interoperability of Metadata. *AI Magazine*, *24*(3), 75. https://doi.org/10.1609/aimag.v24i3.1720

European Commission, Directorate General for Research and Innovation. (2021). *European Research Area policy agenda: Overview of actions for the period 2022-2024.* [Policy]. Publications Office. https://doi.org/10.2777/52110

Fanini, B., Ferdani, D., Demetrescu, E., Berto, S., & d'Annibale, E. (2021). ATON: An Open-Source Framework for Creating Immersive, Collaborative and Liquid Web-Apps for Cultural Heritage. *Applied Sciences*, *11*(22), 11062. https://doi.org/10.3390/app112211062

Genovese, G., & Montanari, R. (2024). *Stakeholder Analysis* (Version 1.0). University of Suor Orsola Benincasa. https://doi.org/10.5281/zenodo.13146893

*Piano Nazionale di Ripresa e Resilienza*. (2021). Italy. https://www.italiadomani.gov.it/content/sogei-ng/it/it/strumenti/documenti/archivio-documenti/piano-nazionale-di-ripresa-e-resilienza.html

Mensa, E., Pecora, A. E., Fulfaro, C., Pizzo, A., Ferraris, E., Bottino, A., & Damiano, R. (2026). From a True Story: Leveraging Museum Catalogue Data for LLM-Driven Narrative Generation. *ACM Transactions on Intelligent Systems and Technologies*. https://doi.org/10.1145/3842754

Mensa, E., Fulfaro, C., Fubini, F., Bottino, A., Antonino, R., Ferraris, E., & Damiano, R. (2025). '"There was a scribe, a priest and a thief"'. Testing the potential of language models for the creation of curatorial narratives in an archaeological museum. *Proceedings of DH2025 - Digital Heritage International Congress 2025*. Digital Heritage 2025 (DH 2025), Florence, Italy. https://doi.org/10.2312/dh.20253118

Reichheld, F. F. (2003). The one number you need to grow. *Harvard Business Review*, *81*(12), 46–54.

Schrepp, M., Hinderks, A., & Thomaschewski, J. (2017). Design and Evaluation of a Short Version of the User Experience Questionnaire (UEQ-S). *International Journal of Interactive Multimedia and Artificial Intelligence*, *4*(6), 103–108. https://doi.org/10.9781/ijimai.2017.09.001

UNESCO. (2009). *Charter on the Preservation of the Digital Heritage* (Circular Letter CL/3865; p. 5). https://unesdoc.unesco.org/ark:/48223/pf0000179529

UNESCO. (2021). *UNESCO Recommendation on Open Science* (Programme and Meeting Document SC-PCB-SPP/2021/OS/UROS; p. 36). https://doi.org/10.54677/MNMH8546

Wilkinson, M. D., Dumontier, M., Aalbersberg, I. J., Appleton, G., Axton, M., Baak, A., Blomberg, N., Boiten, J.-W., da Silva Santos, L. B., Bourne, P. E., Bouwman, J., Brookes, A. J., Clark, T., Crosas, M., Dillo, I., Dumon, O., Edmunds, S., Evelo, C. T., Finkers, R., … Mons, B. (2016). The FAIR Guiding Principles for scientific data management and stewardship. *Scientific Data*, *3*, 160018. https://doi.org/10.1038/sdata.2016.18

# Appendix 1: Spoke 4’s Deliverables

The table in this appendix contains all the 18 deliverables produced within Spoke 4, with their title, work package (WP), version, date of publication, the DOI URL to Zenodo.

| # | Title | WP | Version | Date | DOI URL to Zenodo |
|---|---|---|---|---|---|
| D1 | Spoke coordination guide | WP1 | V1.1 | 2023-05-27 | https://doi.org/10.5281/zenodo.7977471 |
| D2 | Data Management Plan: First version | WP1 | V1.0 | 2023-05-31 | https://doi.org/10.5281/zenodo.7977103 |
| D3 | Meta-analysis of the cultural context | WP2 | V1.0 | 2023-08-31 | https://doi.org/10.5281/zenodo.8305684 |
| D4 | Meta-analysis of the technological context | WP3 | V1.0 | 2023-08-31 | https://doi.org/10.5281/zenodo.8319840 |
| D5 | Preliminary analysis of pilot studies: Status and requirements | WP4 | V1.0 | 2023-11-30 | https://doi.org/10.5281/zenodo.10231012 |
| D6 | Involvement and Exploitation Strategy | WP5 | V1.0 | 2023-11-30 | https://doi.org/10.5281/zenodo.10231205 |
| D7 | Guidelines and best practices for technology-aided narratives for museums and art collections | WP2 | V1.0 | 2024-03-27 | https://doi.org/10.5281/zenodo.10888767 |
| D8 | Data Management Plan: Second version | WP1 | V1.0 | 2024-02-29 | https://doi.org/10.5281/zenodo.10727879 |
| D9 | Implementation of guidelines and best practices for technology-aided narratives via prototypes | WP3 | V1.0 | 2024-11-30 | https://doi.org/10.5281/zenodo.14251585 |
| D10 | Data Management Plan: Final version | WP1 | V1.0 | 2025-02-28 | https://doi.org/10.5281/zenodo.14955816 |
| D11 | Implementation of pilot studies – Type A | WP4 | V1.0 | 2025-09-30 | https://doi.org/10.5281/zenodo.17237860 |
| D12 | Implementation of pilot studies – Type B | WP4 | V1.0 | 2025-09-30 | https://doi.org/10.5281/zenodo.17237906 |
| D13 | Implementation of pilot studies – Type C | WP4 | V1.0 | 2025-09-30 | https://doi.org/10.5281/zenodo.17237965 |
| D14 | Implementation of pilot studies – Type D | WP4 | V1.0 | 2025-09-30 | https://doi.org/10.5281/zenodo.17238011 |
| D15 | Implementation of pilot studies – Type E | WP4 | V1.0 | 2025-09-30 | https://doi.org/10.5281/zenodo.17238050 |

| D16 | Implementation of pilot studies – Type F | WP4 | V1.0 | 2025-09-30 | https://doi.org/10.5281/zenodo.17238078 |
|---|---|---|---|---|---|
| D17 | Meta analysis of implementation of ‘core’ case studies: Identification of technological gaps and further lines of research | WP4 | V1.0 | 2026-02-28 | https://doi.org/10.5281/zenodo.18816266 |
| D18 | Involvement and Exploitation Strategy: Final version | WP5 | V1.0 | 2026-02-28 | https://doi.org/10.5281/zenodo.18816319 |